\documentclass[12pt,aps,prb,preprint]{revtex4}   

\usepackage{amsmath}    
\usepackage{amsfonts}  
\usepackage{amssymb}
\usepackage{graphicx}   

\usepackage[latin1]{inputenc}     
\usepackage[T1]{fontenc}

\usepackage{float}   

\newtheorem{Theoreme}{THEOREM}
\newtheorem{Definition}{Definition}

\draft

\begin{document}

\title{Discerned and Non-Discerned Particles in Classical Mechanics and
Quantum Mechanics Interpretation}
\author{Michel Gondran}
\email{michel.gondran@polytechnique.org} \affiliation{University
Paris Dauphine, Lamsade, Paris, France}
\author{Alexandre Gondran}
 \affiliation{University of Toulouse, ENAC, Toulouse, France}

\begin{abstract}
We introduce into classical mechanics the concept of non-discerned
particles for particles that are identical, non-interacting and
prepared in the same way. The non-discerned particles correspond
to an action and a density which satisfy the statistical
Hamilton-Jacobi equations and allow to explain
 the Gibbs paradox in a simple manner. On the other hand, a discerned particle
corresponds to a particular action that satisfies the local
Hamilton-Jacobi equations. We then study the convergence of
quantum mechanics to classical mechanics when $\hbar\to 0 $ by
considering the convergence for the two cases. These results
provide an argument for a renewed interpretation of quantum
mechanics.
\end{abstract}

 \maketitle

\section{Introduction}
The indiscernability concept, which is very relevant in quantum
and statistical physics, is not well-defined in the literature. In
particular, it is the origin of the Gibbs paradox. Indeed,
\textit{when one calculates the entropy of two mixed gases, the
classical result for distinguishable particles is double the
expected result. If the particles are considered
indistinguishable, the correct result is recovered because of the
indiscernability factor.} This paradox identified by
Gibbs~\cite{Gibbs_1960} in 1889, was solved only by means of
quantum mechanics 35 years later by using the indiscernability
postulate for quantum particles. Indeed it was Einstein who, in
1924, introduced the indiscernability of perfect gas molecules at
the same time as Bose-Einstein statistics. In his homage to
Einstein for the centenary of his birth in 1979, Alfred Kastler
pointed out that~\cite{Kastler_1979}: " \textit{the distinction
between distinguishable and indistinguishable entities and the
difference of statistical behavior between those two types of
entities remains obscure. Boltzmann treated those 'molecules' as
distinguishable entities, which has yielded the so-called
Boltzmann statistics. On the contrary, Planck  implicitly dealt
with the "energy elements" he introduced as indistinguishable
particles, which led to a probabilistic counting of a macroscopic
state different from the Boltzmann one. In 1909, Einstein rightly
criticized this lack of rigor.}" But as noted by Henri
Bacry~\cite{Bacry} p.129, \textit{"the historical progression
could have been very different. Indeed, logically, one could
postulate the non-discernability principle in order to solve the
Gibbs paradox. But this principle can be applied to all the
principles of quantum mechanics or to those of  classical
mechanics.}" This same observation has been made by a large number
of other authors. In 1965 Landé~\cite{Lande_1965} demonstrated
that this indiscernability postulate of classical particles is
sufficient and necessary in order to explain why entropy vanished.
In 1977, Leinaas and Myrheim~\cite{Leinaas_1976} used it for the
foundation of their identical classical and quantum particles
theory. Moreover, as noted by Greiner, in addition to the Gibbs
paradox, several cases where it is needed to consider
indistinguishable particles in classical mechanics and
distinguishable particles in quantum mechanics can be found
~\cite{Greiner_1999} p.134 : "\textit{Hence, the Gibbs factor $
\frac{1}{N!}$ is indeed the correct recipe  for avoiding the Gibbs
paradox. From now on we will therefore always take into account
the Gibbs correction factor for indistinguishable states when we
count the microstates. However, we want to emphasize that this
factor is no more than a recipe to avoid the contradictions of
classical statistical mechanics. In the case of distinguishable
objects (e.g., atoms which are localized at certain grid points),
the Gibbs factor must \textbf{not} be added. In classical theory
the particles remain distinguishable. We will meet this
inconsistency more frequently in classical statistical
mechanics.}" But nowadays, most of the textbooks contain
definitions such as :"\textit{in classical mechanics, two
particles in a system are always distinguishable" and "in quantum
mechanics, two particles are always
indistinguishable}"~\cite{Dalibard_2005} p.328-329, do not answer
the concrete problems exposed by Greiner in both classical
statistics mechanics and quantum statistical mechanics. In this
article we propose an accurate definition of both discernability
and indiscernability in classical mechanics and a way to avoid
ambiguities and paradoxes. These definitions yield an
understandable interpretation both of the action in classical
mechanics and the wave function in quantum mechanics. We only
consider the case of a single particle or a system of identical
particles without interactions and prepared in the same way. The
case of identical particles with interactions will be presented in
a future paper \cite{Gondran_2010b}. In paragraph 2, we introduce
the discerned and non-discerned particles concepts in classical
mechanics through the Hamilton-Jacobi equations. In the following
paragraphs, we study the convergence of quantum mechanics to
classical mechanics when $\hbar$ tends to 0 by considering two
cases: the first corresponds to the convergence to non-discerned
classical particles, and the second corresponds to the convergence
to a classical discerned particle. Based on these convergences, we
propose an updated interpretation
 of quantum mechanics.

\section{Discerned and non-discerned particles in classical mechanics}
Let us consider in classical mechanics a system of identical
particles without interactions.
\begin{Definition}\label{defparticulesdiscernedenmc}- A classical particle is \textbf{potentially discerned} if its initial
position $\textbf{x}_o$ and its initial velocity $\textbf{v}_0$
are known.
\end{Definition}
Let us note that there is an abuse of language when one talks
about a classical particle. One should rather speak of a particle
that is studied in the framework of classical mechanics.

We now consider a particle within a beam of classical identical
particles such as electronic, atomic or molecular beams ($CO_2$ or
$C_{60}$). For such particle, one only knows, initially, the
probability density $\rho _{0}\left( \mathbf{x}\right)$ and the
velocity field $\mathbf{v}_0(\textbf{x})$ through the action
$S_0(\textbf{x})$; this action is known to within a constant from
the equation $\mathbf{v}_0(\textbf{x})= \frac{\nabla
S_0(\textbf{x})}{m}$ where $m$ is the particle mass. This yields
the following definition :

\begin{Definition}\label{defparticulesindiscernedenmc}- A classical particle,
of which at initial time only the density of its initial position
$ \rho _{0}\left( \textbf{x}\right)$ and initial action
$S_0(\textbf{x}) $, is referred to as \textbf{potentially}
\textbf{non-discerned}.
\end{Definition}
This notion is intrinsic to a particle. It gives the initial
conditions, which means the way it has been prepared. Therefore,
it is an indiscernability on the initial particle position. It
doesn't depend on the observer but on the effective modeling
scale of the phenomenon.

\bigskip

In this article, we are only interested in the case where the
\textbf{N particles are prepared in the same way} with the same
initial density $ \rho _{0}\left( \textbf{x}\right)$ and the same
initial action $S_0(\textbf{x})$ evolving in the same potential
$V(x)$ and which can have independent behaviors. It is the case of
classical identical particles \textbf{without interactions} and
prepared in the same way, such as $C_{60}$ or neutral molecules.
It is still the case for instance for electrons prepared in the
same way, and although they are able to interact with each other,
they will have independent behaviors because they are generated
one by one in the system. The general case of interacting
identical particles which are not prepared in the same way will be
presented in a future article~\cite{Gondran_2010b}.

\begin{Definition}\label{defNparticulesindiscernedenmc}- N indentical particles, prepared in the same way,
with the same initial density $ \rho _{0}\left(
\textbf{x}\right)$, the same initial action $S_0(\textbf{x})$, and
evolving in the same potential $V(\textbf{x})$ are called
\textbf{non-discerned}.
\end{Definition}
We have named those particles non-discerned and not
indistinguishable because, if their initial positions are known,
their trajectories will be known as well. Nevertheless, when one
counts them, they will have the same properties as the
indistinguishable ones. Thus, if the initial density $\rho
_{0}\left( \textbf{x}\right)$ is given, and one randomly chooses
$N$ particles, the N! permutations are strictly equivalent and do
not correspond to the same configuration as for indistinguishable
particles. This means that if $X$ is the coordinate space of a
non-discerned particle, the true configuration space of $N$
non-discerned particles is not $X^N$ but rather $ X^N / S_N $
where $S_N$ is the symmetric group.

\subsection{Non-discerned particles and statistical Hamilton-Jacobi equations}

For non-discerned particles, we have the following theorem:

\begin{Theoreme}\label{th:eqstatHJ}- The probability density
$\rho \left( \mathbf{x},t\right)$ and the action $S\left(
\mathbf{x,}t\right)$ of classical particles prepared in the same
way, with initial density $\rho_0( \textbf{x})$, with the same
initial action $S_0(\textbf{x})$, and evolving in the same
potential $V(\textbf{x})$, are solutions to the
\textbf{statistical Hamilton-Jacobi equations}:
\begin{eqnarray}\label{eq:statHJ1b}
\frac{\partial S\left(\textbf{x},t\right) }{\partial
t}+\frac{1}{2m}(\nabla S(\textbf{x},t) )^{2}+V(\textbf{x})=0
&\text{\ \ }&
\forall \left( \textbf{x},t\right) \in
\mathbb{R}^{3}\times \mathbb{R}^{+}
\\
\label{eq:statHJ2b}
S(\textbf{x},0)=S_{0}(\textbf{x})
&\text{\ \ }&
\forall\textbf{x}\in \mathbb{R}^{3}.
\\
\label{eq:statHJ3b}
\frac{\partial \mathcal{\rho }\left(\textbf{x},t\right) }{\partial
t}+ div \left( \rho \left( \textbf{x},t\right) \frac{\nabla
S\left( \textbf{x},t\right) }{m}\right) =0
&\text{\ \ }&
\forall \left( \textbf{x},t\right) \in \mathbb{R}^{3}\times
\mathbb{R}^{+}
\\
\label{eq:statHJ4b}
\rho(\mathbf{x},0)=\rho_{0}(\mathbf{x})
&\text{\ \ }&
\forall
\mathbf{x}\in \mathbb{R}^{3}.
\end{eqnarray}

\end{Theoreme}
\bigskip
Let us recall that the velocity field is
$\textbf{v}(\textbf{x},t)= \frac{\nabla S\left(
\textbf{x},t\right) }{m}$ and that the Hamilton-Jacobi equation
(\ref{eq:statHJ1b}) is not coupled to the continuity equation
(\ref{eq:statHJ3b}). The difference between discerned particles
and non-discerned particles thus explains why the "recipes"
proposed in some classical statistical mechanics books are useful.
But as has been demonstrated above, it is not a principle which
can be added. The nature of the discernability of the particles
depends strongly on the experimental conditions determined by the
modeling scale.


\subsection{Discerned particles and local Hamilton-Jacobi equations}
One can ask if it is possible to define an action for a
potentially discerned particle in a potential field
$V(\textbf{x})$?  Such an action should depend only on the
starting point $\textbf{x}_0$, the initial velocity $\textbf{v}_0$
and the potential $V(\textbf{x})$.

\begin{Theoreme}\label{th:actionponctuelle}- If $\xi(t)$
is the classical trajectory in the field $V(\textbf{x})$ of a
particle with the initial position $\textbf{x}_0$ and with initial
velocity $\textbf{v}_0$, then the function
\begin{equation}\label{eq:soleqHJponctuelle}
S^{\xi}\left( \mathbf{x},t\right)= m \frac{d \xi(t)}{dt} \cdot
\textbf{x} + g(t)
\end{equation}
where $\frac{d g(t)}{dt}= -\frac{1}{2}m (\frac{d \xi(t)}{dt})^2-
V(\xi(t)) - m \frac{d^2 \xi(t)}{dt^2} \cdot \xi(t)$, is called
\textbf{local action}, and is solution to local Hamilton-Jacobi
equations.
\begin{eqnarray}\label{eq:statHJponctuelle1b}
\frac{\partial S^{\xi}\left(\textbf{x},t\right) }{\partial
t}|_{\textbf{x}=\xi(t)}+\frac{1}{2m}(\nabla S^{\xi}(\textbf{x},t)
)^{2}|_{\textbf{x}=\xi(t)}+V(\textbf{x})|_{\textbf{x}=\xi(t)}=0
&\text{ \ \ }&
\forall t \in \mathbb{R}^{+}
\\
\label{eq:statHJponctuelle1c}
\frac{d\xi(t)}{dt}=\frac{\nabla S^{\xi}(\xi(t),t)}{m}
&\text{ \ \ }&
\forall t \in \mathbb{R}^{+}
\\
\label{eq:statHJponctuelle1d}
S^{\xi}(\textbf{x},0)= m \textbf{v}_0 \textbf{x}
\text{ \ et \ }\xi(0)=\textbf{x}_0.
\end{eqnarray}
\end{Theoreme}
The local action satisfies the Hamilton-Jacobi equations only
along the trajectory $\xi(t)$. The introduction of such an action
linked to a trajectory appears as strange and devoid of any
effective interest other than a theoretical one by proposing a
framework for defining discerned particles. This action will take
on a meaning in paragraph 4 when we show that it corresponds to
the convergence of coherent state when $\hbar$ tends to 0. We have
defined two kinds of actions, a global one $S(\mathbf{x},t)$ and a
local one $ S^{\xi}\left(\textbf{x},t\right)$. The global action
$S(\mathbf{x},t)$ is a field defined for all $\textbf{x}$
independently of the starting point $\textbf{x}_0$. But the local
one $ S^{\xi}\left(\textbf{x},t\right)$ depends on the trajectory
$\xi(t) $ and the starting point $\textbf{x}_0$. \textbf{The least
action principle is valid only for the global action and not for
the local one.} This difference provides an answer to the doubts
emitted by some physicists bothered by the use of the least action
principle. In particular, Henri Poincaré who wrote in "La science
et l'hypoth\`ese":~\cite{Poincare_1902}

\textit{"The statement of the least action principle is somehow
shocking for the mind. To move from one point to another, a
material molecule, removed from the action of any force, but
subject to mmoving on a surface, will move through the geodesic
line, which means the shortest path. This molecule seems to know
the point one wishes to guide it to, to predict the time it needs
to reach it by choosing one path or another and to choose the most
suitable one. The statement thus presents the particle as a free
and animated being. It is clear that it would be better to replace
it with a less shocking statement and where, as the philosophers
would say, the final causes would not seem to be taking the place
of efficient causes."} This paradox can be solved if one remarks
that the least action principle can only be applied to a global
action and not to a local one because this former one depends on
the starting or final point.

\bigskip

\section{Convergence to non-discerned particles when $\hbar \to 0$.}

Let us consider the wave function solution to the Schr\"odinger
equation $\Psi(\textbf{x},t)$:
\begin{eqnarray}\label{eq:schrodinger1}
i\hslash \frac{\partial \Psi }{\partial t}=\mathcal{-}\frac{\hslash ^{2}}{2m}%
\triangle \Psi +V(\mathbf{x})\Psi
&\text{ \ \ }&
\forall (\mathbf{x},t)\in \mathbb{R}%
^{3}\times \mathbb{R}^{+}
\\
\label{eq:schrodinger2}
\Psi (\mathbf{x},0)=\Psi_{0}(\mathbf{x})
&\text{ \ \ }&
\forall \mathbf{x}\in \mathbb{R}^{3}.
\end{eqnarray}
With the variable change $ \Psi
(\mathbf{x},t)=\sqrt{\rho^{\hbar}(\mathbf{x},t)} \exp(i
\frac{S^{\hbar}(\textbf{x},t)}{\hbar})$, the density
$\rho^{\hbar}(\mathbf{x},t)$ and the action
$S^{\hbar}(\textbf{x},t)$ are on the parameter $\hbar$. The
Schrödinger equation may be divided into Madelung
equations~\cite{Madelung_1926} (1926) which correspond to:
\begin{eqnarray}\label{eq:Madelung1}
\frac{\partial S^{\hbar}(\mathbf{x},t)}{\partial t}+\frac{1}{2m}
(\nabla S^{\hbar}(\mathbf{x},t))^2 +
V(\mathbf{x})-\frac{\hbar^2}{2m}\frac{\triangle
\sqrt{\rho^{\hbar}(\mathbf{x},t)}}{\sqrt{\rho^{\hbar}(\mathbf{x},t)}}=0
&\text{ \ \ }&
\forall (\mathbf{x},t)\in \mathbb{R}%
^{3}\times \mathbb{R}^{+}
\\
\label{eq:Madelung2}
\frac{\partial \rho^{\hbar}(\mathbf{x},t)}{\partial t}+ \nabla
\cdot (\rho^{\hbar}(\mathbf{x},t)
\frac{\nabla S^{\hbar}(\mathbf{x},t)}{m})=0
&\text{ \ \ }&
\forall (\mathbf{x},t)\in \mathbb{R}%
^{3}\times \mathbb{R}^{+}
\end{eqnarray}
with initial conditions
\begin{equation}\label{eq:Madelung3}
\rho^{\hbar}(\mathbf{x},0)=\rho^{\hbar}_{0}(\mathbf{x}) \text{ \ et \ }
S^{\hbar}(\mathbf{x},0)=S^{\hbar}_{0}(\mathbf{x}) \qquad
\forall \mathbf{x}\in \mathbb{R}^{3}.
\end{equation}

In the two following paragraphs we study the convergence of the
density $\rho^{\hbar}(\mathbf{x},t)$ and the action
$S^{\hbar}(\textbf{x},t)$ in the Madelung equations when $\hbar$
tends to 0. It is subtle and remains a difficult problem. For this
reason, we only consider two typical cases, for which analytical
solutions exist. The difference between the two examples is the
fact that \textit{the initial conditions are not the same due to a
different preparation of the particles and initial conditions for
the potential when $ \hbar$ tends to 0}.

\begin{Definition}\label{defdensiteinitstat}- A quantum system is \textbf{non-discerned semi-classically} if it satisfies the two following conditions

- its initial probability density $\rho^{\hbar}_{0}(\mathbf{x})$
and its initial action $S^{\hbar}_{0}(\mathbf{x})$ converge
respectively, to regular functions $\rho_{0}(\mathbf{x})$ and
$S_{0}(\mathbf{x})$ not depending on $\hbar$ when $\hbar\to 0$.

- its interaction with the potential field $V(\textbf{x})$ can be
described classically. The simplest case corresponds to particles
in vacuum with only geometric constraints. For instance, Young's
slits interference experiment, or a single particle in a box
($V(\textbf{x})= 0$ or $V(\textbf{x})=+ \infty$).
\end{Definition}
As previously described, this is the case of a set of
non-interacting particles prepared in the same way: free particles
beam in a linear potential, electronic or $C_{60}$ beam in the
Young's slits diffraction, atomic beam in Stern and Gerlach
experiment.

\subsection{Convergence to statistical Hamilton-Jacobi equations}

If we consider the system with classical initial conditions
\begin{equation}\label{eq:densiteinit}
\rho^{\hbar}_{0}(\mathbf{x})=\rho_{0}(\mathbf{x})=( 2\pi \sigma
_{0}^{2}) ^{-\frac{3}{2}}e^{-\frac{( \textbf{x}-\zeta_{0})
^{2}}{2\sigma _{0}^{2}}} \qquad \text{and} \qquad
S^{\hbar}_{0}(\mathbf{x})= S_{0}(\mathbf{x})= m
\textbf{v}_{0}\cdot \textbf{x}.
\end{equation}
in a linear potential field $V(\textbf{x})=-
\textbf{K}\cdot\textbf{x}$. The density
$\rho^{\hbar}(\textbf{x},t)$ and the action
$S^{\hbar}(\textbf{x},t)$, solutions to the Madelung equations
(\ref{eq:Madelung1})(\ref{eq:Madelung2})(\ref{eq:Madelung3}) with
the initial condition (\ref{eq:densiteinit}), are respectively
equal to~\cite{CohenTannoudji_1977}~:
\begin{equation}\label{eq:densite}
\rho^{\hbar}(\textbf{x},t)=( 2\pi \sigma_{\hbar} ^{2}( t))
^{-\frac{3}{2}}e^{- \frac{\left(
\textbf{x}-\zeta_{0}-\textbf{v}_{0}t-\textbf{K} \frac{t^{2}}{2
m}\right) ^{2}}{2\sigma_{\hbar} ^{2}( t) }}
\end{equation}
\begin{eqnarray}
S^{\hbar}(\textbf{x},t)=&-&\frac{3 \hbar }{2}tg^{-1}( \hbar t/2m
\sigma _{0}^{2}) - \frac{1}{2}m\textbf{v}_{0}^{2}t+ m
\textbf{v}_{0}\cdot \textbf{x}+\textbf{K}\cdot\textbf{x} t \nonumber\\
&-&
\frac{1}{2} \textbf{K}\cdot\textbf{v}_{0} t^{2} -
\frac{\textbf{K}^2 t^3}{6 m}
+\frac{\left(
\textbf{x}-\zeta_{0}-\textbf{v}_{0}t -\textbf{K} \frac{t^{2}}{2
m}\right) ^{2}\hbar ^{2}t}{8m\sigma _{0}^{2}\sigma_{\hbar}
^{2}\left( t\right) }
\end{eqnarray}
with
\begin{equation}\label{eq:sigmah}
\sigma_{\hbar} \left( t\right) =\sigma _{0}\left( 1+\left( \hbar
t/2m\sigma _{0}^{2}\right) ^{2}\right) ^{\frac{1}{2}}.
\end{equation}

The constants $\sigma_0$, $\textbf{v}_0$, $\zeta_0$ and
$\textbf{K}$ are given and independent of $\hbar$; $\sigma_0$ for
example corresponds to the hole width for preparing the particle
beam.

When $\hbar\to 0$, $\sigma_{\hbar} \left( t\right)$ converges to
$\sigma_0$ and one gets the following theorem :

\begin{Theoreme}\label{r-th1}-When $\hbar\to 0$,
the density $\rho^{\hbar}(\textbf{x},t)$ and the action
$S^{\hbar}(\textbf{x},t)$ converge to
\begin{eqnarray}
&&\rho(\textbf{x},t)=( 2\pi \sigma_{0} ^{2}) ^{-\frac{3}{2}}e^{-%
\frac{\left( \textbf{x}-\zeta_{0}-\textbf{v}_{0}t -\textbf{K}
\frac{t^{2}}{2 m}\right) ^{2}}{2\sigma_{0} ^{2} }} \\
&\text{and  }&
S(\textbf{x},t)= - \frac{1}{2}m \textbf{v}_{0}^{2}t+m
\textbf{v}_{0}\cdot\textbf{x}+\textbf{K}\cdot\textbf{x} t -
\frac{1}{2} \textbf{K}\cdot\textbf{v}_{0} t^{2} -
\frac{\textbf{K}^2 t^3}{6 m}.
\end{eqnarray}
which are solutions to statistical Hamilton-Jacobi equations
(\ref{eq:statHJ1b})(\ref{eq:statHJ2b})(\ref{eq:statHJ3b})(\ref{eq:statHJ4b}).
\end{Theoreme}

Thus, when $\hbar\to 0$, for \textit{semi-classical non-discerned
particles}, the probability density $\rho^{\hbar}(\textbf{x},t)$
of the wave function tends to the probability density of a
statistical set of classical particles $\rho(\textbf{x},t)$. We
conjecture that this result in the case of a linear potential
field can be generalized to \textit{semi-classically discerned
particles} for other potentials.

\bigskip

\textbf{CONJECTURE} - \textit{For semi-classically non-discerned
particles, when $\hbar\to 0$, for all $\textbf{x}$ and $t$
bounded, the density $\rho^{\hbar}(\textbf{x},t)$ and the action
$S^{\hbar}(\textbf{x},t)$, which are solutions to Madelung
equations}(\ref{eq:Madelung1})(\ref{eq:Madelung2})(\ref{eq:Madelung3}),
\textit{converge to $\rho(\textbf{x},t)$ et $S(\textbf{x},t)$,
which are solutions to statistical Hamilton-Jacobi equations.}
(\ref{eq:statHJ1b})(\ref{eq:statHJ2b})(\ref{eq:statHJ3b})(\ref{eq:statHJ4b}).

\bigskip
This conjecture is verified for the convergence of the density
$\rho^{\hbar}(\textbf{x},t)$ with an explicit calculation for the
Stern-Gerlach experiment~\cite{Gondran_2005a}, for the EPR
one~\cite{Gondran_2008}, and by numerical simulation for the
Young's slits experiment~\cite{Gondran_2005,Gondran_2010}.

\subsection{De Broglie-Bohm quantum trajectories}

Those last convergence examples show that for
\textit{semi-classically non-discerned particles}, the Madelung
equations converge to statistical Hamilton-Jacobi equations. The
uncertainty of the position of a quantum particle corresponds in
that case to an uncertainty of the position of a classical
particle, only whose initial density has been defined. \textbf{In
classical mechanics, this uncertainty is removed by giving the
initial position of the particle. It would be illogical not to do
the same in quantum mechanics.}

\bigskip
We assume that for \textit{semi-classically non-discerned
particles}, a quantum particle is not well described by its wave
function. It is therefore necessary to add its initial position
and it becomes natural to introduce the so-called de Broglie-Bohm
trajectories. In this interpretation, its velocity is given by
\cite{deBroglie_1927,Bohm_52}:
\begin{equation}\label{eq:vitessequantique}
\textbf{v}^{\hbar}(\textbf{x},t) = \frac{1}{m}\nabla
S^{\hbar}(\textbf{x},t)
\end{equation}
or by the alternative form \cite{Bohm_93, Holland_93, Holland_99,
GondranMA}:
\begin{equation}\label{eq:vitessequantiqueavecspin}
\textbf{v}^{\hbar}(\textbf{x},t) = \frac{1}{m}\nabla
S^{\hbar}(\textbf{x},t) + \frac{\hbar}{2 m}\nabla
\ln\rho^{\hbar}(\textbf{x},t) \times \textbf{k},
\end{equation}
where $\textbf{k}$ is the unit vector parallel to the particle
spin vector.

This spin current $\frac{\hbar}{2 m}\nabla
\rho^{\hbar}(\textbf{x},t) \times \textbf{k}$ corresponds to
Gordon's current when one changes from the Dirac equation to the
Pauli equation and subsequently to the Schrodinger
equation\cite{Holland_99}. This current is very important because
it allows us to return to quantum mechanics on small scales, in
particular in relation to Compton's wavelength, as in the Foldy
and Wouthuysen transformation \cite{Foldy}.

We have the following classical property: if a system of particles
with initial density $\rho_{0}(\mathbf{x})$ has de
Broglie-Bohm-like trajectories defined by the velocity field
$\textbf{v}^{\hbar}(\textbf{x},t)$ from equations
(\ref{eq:vitessequantique}) or
(\ref{eq:vitessequantiqueavecspin}), then the probability density
of those particles at time $t$ is equal to
$\rho^\hbar(\textbf{x},t)$, the square of the wave function
magnitude. In the case of semi-classical non-discerned particles,
this shows that the Broglie-Bohm interpretation reproduces the
predictions of standard quantum mechanics.

In one dimension, for the initial particle position $x_0=\zeta_0 +
\eta_0 $ with initial velocity $v_0$, in a linear potential
$V(x)=- Kx$ and with velocity (\ref{eq:vitessequantique}), one
recovers the Broglie-Bohm trajectory : $\xi_{\hbar}(t)= \zeta_0 +
v_0 t - K \frac{t^2}{2 m}+\eta_0 \frac{\sigma_{\hbar} \left(
t\right)}{\sigma_0}$ which converges to the classical trajectory
$\xi(t)= \zeta_0 +\eta_0 + v_0 t - K \frac{t^2}{2 m}$ when
$\hbar\to 0$.

In three dimensions, for a particle initial position such as
$\textbf{x}_0=\zeta_0 + \eta_0 $ with an initial velocity
$\textbf{v}_0$, in a linear potential $V(\textbf{x})=- K x_3$ and
with the velocity (\ref{eq:vitessequantiqueavecspin}), we have the
Bohm-Broglie trajectory\cite{Gondran_2005} :
$\xi_{0,1}^{\hbar}(t)= \zeta_{0,1} + v_{0,1} t
+\sqrt{\eta^2_{0,1}+ \eta^2_{0,2}} \frac{\sigma_{\hbar} \left(
t\right)}{\sigma_0} \cos \varphi(t)$, $\xi_{2}^{\hbar}(t)=
\zeta_{0,2} + v_{0,2} t + \sqrt{\eta^2_{0,1}+ \eta^2_{0,2}}
\frac{\sigma_{\hbar} \left( t\right)}{\sigma_0} \sin \varphi(t)$,
$\xi_{3}^{\hbar}(t)= \zeta_{0,3} + v_{0,3} t - K \frac{t^2}{2
m}+\eta_{0,3} \frac{\sigma_{\hbar} \left( t\right)}{\sigma_0}$
avec $ \varphi(t)= \arctan (\frac{\eta_{0,1}}{\eta_{0,2}}) -
\arctan (\frac{\hbar t }{2 m \sigma^2_0})$, which converges to the
classical trajectory $\xi_{0,1}(t)= \zeta_{0,1} +\eta_{0,1} +
v_{0,1} t$ , $\xi_{0,2}(t)= \zeta_{0,2} +\eta_{0,2} + v_{0,2} t$,
$\xi_{0,3}(t)= \zeta_{0,3} +\eta_{0,3} + v_{0,3} t - K
\frac{t^2}{2 m}$ when $\hbar\to 0$.

Generally, when $\hbar\to 0$, one deduces from conjecture that
$\textbf{v}^{\hbar}(\textbf{x},t)$ given from equations
(\ref{eq:vitessequantique}) or (\ref{eq:vitessequantiqueavecspin})
converge to the classical velocity $\textbf{v}(\textbf{x},t) =
\frac{1}{m}\nabla S(\textbf{x},t)$. This leads to the fact that
the Broglie-Bohm trajectories converge to the classical ones. We
verify this conjecture with an explicit calculation for the
Stern-Gerlach experiment~\cite{Gondran_2005a} and by numerical
simulation for the Young's slits
experiment~\cite{Gondran_2005,Gondran_2010}.

\section{Convergence to discerned particles when $\hbar \to 0$.}

\begin{Definition}\label{defdensiteinitponct}- A quantum system is \textbf{discerned semi-classically} if it
satisfies the two conditions

- its initial probability density $\rho^{\hbar}_{0}(\mathbf{x})$
and its initial action $S^{\hbar}_{0}(\mathbf{x})$ converge
respectively, when $\hbar\to 0$, to a Dirac distribution and an
action  $S_{0}(\mathbf{x})$ not depending on $\hbar$.

- its interaction with the potential field $V(\textbf{x})$ can be
described classically.
\end{Definition}
This situation occurs when the wave packet corresponds to a
quasi-classical coherent state which were introduced in 1926 by
Schr\"odinger~\cite{Schrodinger_26}, and is of great importance in
quantum optics since Glauber~\cite{Glauber1965}(1965). They have
three properties: their gravity center follows a classical
trajectory; they verify a Heisenberg equality and not an
inequality; the wave packet shape doesn't change during motion (or
at least it recovers its shape after a cycle). It still occurs
when the wave packet corresponds to the periodic trajectories of a
non-dispersive wave packet, which are eigenvectors of the Floquet
operator. For the hydrogen atom, the existence of a localized wave
packet on the classical trajectory (an old dream of
Schr\"odinger's) and which was predicted in 1994 by
Bialynicki-Birula, Kalinski, Eberly, Buchleitner et
Delande~\cite{Bialynicki_1994, Delande_1995, Delande_2002}, has
been discovered recently by Maeda and Gallagher~\cite{Gallagher}
on Rydberg atoms.

\subsection{Convergence of coherent states to the solutions to the local Hamilton-Jacobi equations}

For the two dimensional harmonic oscillator,
$V(\textbf{x})=\frac{1}{2}m \omega^{2}\textbf{x}^{2}$, coherent
states are built~\cite{CohenTannoudji_1977} from the initial wave
function $\Psi_{0}(\textbf{x})$ which corresponds to the density
and initial action:
\begin{equation}\label{eq:densiteactioninith}
\rho^{\hbar}_{0}(\mathbf{x})= ( 2\pi \sigma _{\hbar}^{2}) ^{-1}
e^{-\frac{( \textbf{x}-\textbf{x}_{0}) ^{2}}{2\sigma
_{\hbar}^{2}}} \text{ \ and \ }
S_{0}(\mathbf{x})=S^{\hbar}_{0}(\mathbf{x})= m \textbf{v}_{0}\cdot
\textbf{x}
\end{equation}
with $ \sigma_\hbar=\sqrt{\frac{\hbar}{2 m \omega}}$. Here,
$\textbf{v}_0$ and $\textbf{x}_0$ are still constant vectors and
independent from $\hbar$, but $\sigma_\hbar$ will tend to $0$ as
$\hbar$.

For this harmonic oscillator, the density
$\rho^{\hbar}(\textbf{x},t)$ and the action
$S^{\hbar}(\textbf{x},t)$,solutions to Madelung equations
(\ref{eq:Madelung1})(\ref{eq:Madelung2})(\ref{eq:Madelung3}) with
initial conditions (\ref{eq:densiteactioninith}), are equal to
~\cite{CohenTannoudji_1977}:
\begin{equation}
\rho^{\hbar}(\textbf{x},t)=\left( 2\pi \sigma_{\hbar} ^{2} \right)
^{-1}e^{- \frac{( \textbf{x}-\xi(t)) ^{2}}{2\sigma_{\hbar} ^{2}
}}\text{ \ and \ }S^{\hbar}(\textbf{x},t)=  + m \frac{d\xi
(t)}{dt}\cdot \textbf{x} + g(t) - \hbar\omega t
\end{equation}
where $\xi(t)$ is the trajectory of a classical particle evolving
in the potential $V(\textbf{x})=\frac{1}{2} m \omega^{2}
\textbf{x}^2 $, with $\textbf{x}_0$ and $\textbf{v}_0$ as initial
position and velocity where $g(t)=\int _0 ^t ( -\frac{1}{2} m
(\frac{d\xi (s)}{ds})^{2} + \frac{1}{2} m \omega^{2} \xi(s)^2)
ds$. Because we have $2 V(\xi(s))=m \frac{d^2 \xi(s)}{ds^2} \cdot
\xi(s) $, it yields the following theorem :

\begin{Theoreme}\label{t-convergenceparticulediscerne}- When $\hbar\to 0$,
for all $\textbf{x}$ and $t$ bounded, the density
$\rho^{\hbar}(\textbf{x},t)$ and the action
$S^{\hbar}(\textbf{x},t)$ converge respectively to
$\rho^{\xi}(\textbf{x},t)=\delta( \textbf{x}- \xi(t))$ and
$S^{\xi}(\textbf{x},t)= m \frac{d\xi (t)}{dt}\cdot\textbf{v}_{0} +
g(t)$ where $S^{\xi}(\textbf{x},t)$ and the trajectory $\xi(t) $
are solutions to the local Hamilton-Jacobi equations
(\ref{eq:statHJponctuelle1b})(\ref{eq:statHJponctuelle1c})(\ref{eq:statHJponctuelle1d}).
\end{Theoreme}
Therefore, the kinematic of the wave packet converges to the
single harmonic oscillator described by $\xi(t)$. Because this
classical particle is completely defined by its initial conditions
$\textbf{x}_0$ and $\textbf{v}_0$, it can be considered as a
discerned particle.

When $\hbar\to 0$, for all $\textbf{x}$ and $t$ bounded, the
"quantum potential"
$Q^{\hbar}(\textbf{x},t)=-\frac{\hbar^2}{2m}\frac{\triangle
\sqrt{\rho}}{\sqrt{\rho}}= \hbar \omega - \frac{1}{2} m \omega^{2}
(\textbf{x} -\xi(t))^{2}$ tends to $Q(\textbf{x},t)= - \frac{1}{2}
m \omega^{2} (\textbf{x} -\xi(t))^{2}$. It is then zero on the
trajectory ($\textbf{x}=\xi(t)$).

\bigskip

More generally, let us consider semi-classically non-discerned
particles where $\rho^{\hbar}(\textbf{x},t)$ converge to Dirac
distribution $\rho^{\xi}(\textbf{x},t)=\delta( \textbf{x}-
\xi(t))$. Mathematically, one needs, as proposed by Kazandjian
\cite{Kazandjian_2006}, to study the  convergence of Madelung
equations in the least square approach. This yields:
\begin{equation}\label{eq:Madelungmoindrecarre}
\int \rho_{\hbar}(\textbf{x},t)[\frac{\partial
S^{\hbar}(\textbf{x},t)}{\partial t}+\frac{1}{2m} (\nabla
S^{\hbar}(\textbf{x},t))^2 + V(\mathbf{x})+
Q^{\hbar}(\textbf{x},t)]^{2} d\textbf{x}=0.
\end{equation}
For $\hbar\neq 0$, the equation (\ref{eq:Madelungmoindrecarre})
yields the Madelung equations (\ref{eq:Madelung1}); in the limit
$\hbar \rightarrow 0 $, but, $\rho_{\hbar}(\textbf{x},t)$ tends to
0 for all $\textbf{x} \neq \xi(t)$, and converges to a Dirac
distribution centered on $\xi(t)$; In this limit, the equation
(\ref{eq:Madelungmoindrecarre}) can be written as
\begin{equation*}
\frac{\partial S(\xi(t),t)}{\partial t}+\frac{1}{2m} (\nabla
S(\xi(t),t))^2 + V(\xi(t))+ Q(\xi(t),t)=0.
\end{equation*}
And because $Q(\xi(t),t)=0$, this yields
\begin{equation}\label{eq:Madeluugponctuel}
\frac{\partial S(\xi(t),t)}{\partial t}+\frac{1}{2m} (\nabla
S(\xi(t),t))^2 + V(\xi(t))=0
\end{equation}
which corresponds to the local Hamilton-Jacobi equations
(\ref{eq:statHJponctuelle1b}). Thus, in the general case of
semi-classically non-discerned particles, the wave function
kinematics converges to the motion of a discerned classical
particle $\xi(t)$ which is completely defined by its initial
position $\textbf{x}_0$ and its initial velocity $\textbf{v}_0 $.

It is then possible to consider, unlike in the semi-classically
non-discerned case, that the wave function can be seen as a single
quantum particle. The \textit{semi-classically discerned case} is
in agreement with the Copenhagen interpretation of the wave
function, which contains all the information on the particle.

\subsection{Interpretation for the semi-classically discerned particles}

\bigskip

In the \textit{semi-classically discerned case}, the Broglie-Bohm
interpretation is not relevant mathematically, unlike the
semi-classically non-discerned case. Other assumptions are
possible. A natural interpretation is the one proposed by
Schr\"odinger ~\cite{Schrodinger_26} in 1926 for the coherent
states of the harmonic oscillator. In the \textbf{Schr\"odinger
interpretation}, the quantum particle in the semi-classically
discerned case is a spatially extended particle, represented by a
wave packet whose center follows the classical trajectory. For the
coherent states of the harmonic oscillator in two dimensions, the
velocity field (\ref{eq:vitessequantiqueavecspin}) at time $t$ and
at point $\textbf{x}$ is then equal to :
\begin{equation}\label{eq:eqvitessemqosc}
\textbf{v}^{\hbar}(\mathbf{x},t)=\textbf{v}(t) + \Omega \times
(\textbf{x} -\xi(t))
\end{equation}
with $\Omega = \omega \textbf{k}$. They behave as extended
particles which have the same evolution as spinning particles in
two dimensions. But this cannot be generalized easily in three
dimensions. It seems that it is not possible to consider in three
dimensions the particle as a solid in motion. This is the main
difficulty in the Schrödinger interpretation: does the particle
exist within the wave packet? We think that this reality can only
be defined on the scale where the Schrödinger equation is the
effective equation. Some solutions are nevertheless possible on
smaller scales~\cite{Gondran_2001,Gondran_2004}, where the quantum
particle is not represented by a point but is a sort of elastic
string whose gravity center follows the classical trajectory
$\xi(t)$.

\bigskip

Another possible interpretation for the semi-classical discerned
particles is the Bohr model of the atom (1913) found again by de
Broglie~\cite{deBroglie_1924} in 1924 with conditions of resonance
between the wave and the particle. In the Bohr-deBroglie
interpretation, the quantum particle is a point (in relation to
the wave packet size) which follows a trajectory in resonance with
its internal vibration in the wave.

\bigskip
The principle of an interpretation that depends on the particle
preparation conditions is not really new. It has already been
figured out by Einstein and de Broglie. For Louis de Broglie, its
real interpretation was the double solution theory introduced in
1927 in which the pilot-wave is just a low-level product:
 "\textit{I introduced as a 'double solution theory' the idea that
it was necessary to distinguish two different solutions but both
linked to the wave equation, one that I called wave $u$ which was
a real physical wave but not normalizable having a local anomaly
defining the particle and represented by a singularity, the other
one as the Schr\"odinger $\Psi$ of wave, which is normalizable
without singularities and being a probability representation.}"

We consider as interesting L. de Broglie's idea of the existence
of a statistical wave, $\Psi$ and of a soliton wave $u$; however,
it is not a double solution which appears here but a double
interpretation of the wave function according to the initial
conditions.

Einstein's point of view is well summed up in one of his final
papers (1953), "\textit{Elementary reflections concerning the
foundation of quantum mechanics}" in homage to Max Born:

"\textit{The fact that the Schr\"odinger equation associated to
the Born interpretation does not lead to a description of the
"real states" of an individual system, naturally incites one to
find a theory that is not subjected to this limitation.}
\textit{Up to now, the two attempts have in common that they
conserve the Schr\"odinger equation and abandon the Born
interpretation.} \textit{The first one, which marks a de Broglie's
return, was continued by Bohm.}... \textit{The second one, which
aimed to get a "real description" of an individual system and
which might be based on the Schr\"odinger equation, is very late
and is from Schr\"odinger himself. The general idea is briefly the
following : the function $\psi$ represents in itself the reality
and it is not necessary to add Born's statistical
interpretation.}[...] \textit{From previous considerations, it
results that the only acceptable interpretation of the
Schr\"odinger equation is the statistical interpretation given by
Born. Nevertheless, this interpretation doesn't give the 'real
description' of an individual system, it just gives statistical
statements of a set of systems.}"

Thus, it is because de Broglie and Schr\"odinger keep the
Schrödinger equation that Einstein, who considers it as
fundamentally statistical, refused each of their interpretations.

Finally, there exist situations where the \textbf{Broglie-Bohm
interpretation of the Schr\"odinger wave function is probably
wrong}. It is in particular the case of state transitions for a
hydrogen atom. Indeed, since
Delmelt'experiment~\cite{Delmelt_1986} in 1986, the physical
reality of individuals quantum jumps has been fully validated. The
semi-classical approximation, where the interaction with the
potential field can be described classically, is no longer
possible and one must use electromagnetic field quantization since
the exchanges occur photon by photon. Einstein thought that it is
not possible to find an individual deterministic behavior from the
Schrödinger equation. It is the same for Heisenberg who developed
matrix mechanics and the second quantization from this example.

This doesn't mean that one has to renounce to determinism and
realism, but rather that Schr\"odinger's statistical wave function
does not permit, in that case, to discover an individual behavior.

\bigskip

\section{Conclusion}

The introduction into classical mechanics of the concepts of
\textbf{non-discerned particles} and \textbf{discerned particles}
respectively verifying the \textbf{statistical Hamilton-Jacobi
equations} and the \textbf{local Hamilton-Jacobi equations} gives
a simple answer to some paradoxes in classical statistical
mechanics and allows to have a better understanding of the least
action principle.

When one studies the convergence of the Madelung equations when
$\hbar \to 0$, we obtain the following results:

- In the \textit{\textbf{semi-classically non-discerned case}} the
quantum particles converge to classical non-discerned ones,
verifying the statistical Hamilton-Jacobi equations. The wave
function is not sufficient to represent the quantum particles. One
needs to add it the initial positions, as for classical particles,
in order to describe them completely. Thus, \textit{\textbf{the
Broglie-Bohm interpretation is relevant}}.

- In the \textit{\textbf{semi-classically discerned case}} the
quantum particles converge to classical discerned ones, verifying
the local Hamilton-Jacobi equations. \textit{\textbf{The
Broglie-Bohm interpretation is not imperative}} because the wave
function is sufficient to represent the particles as in the
Copenhagen interpretation. However, one can make some realistic
and deterministic assumptions such as the \textbf{Schr\"odinger}
and the \textbf{Bohr-deBroglie interpretations}.

- In the case where \textit{\textbf{the semi-classical
approximation is no longer valid }}, as in the transition states
in the hydrogen atom, the two interpretations are wrong as claimed
by \emph{\textbf{Heisenberg}}. \emph{\textbf{Consequently, Born's
statistical interpretation is the only possible interpretation of
the Schrödinger equation.}} This doesn't mean that it is necessary
to give up to determinism and realism, but rather that the
Schr\"odinger wave function doesn't allow, in that case, to reveal
the individual behavior of a particle. An individual
interpretation needs to use creation and annhilation operators of
quantum Field Theory.

Therefore, as Einstein said, the situation is much more complex
than what de Broglie and Bohm thought.

\end{document}